\documentstyle[12pt,epsfig,equation]{article}

\def\ben{\begin{subequations}}
\def\be{\begin{equation}}
\def\een{\end{subequations}}
\def\ee{\end{equation}}
\def\beq{\begin{eqalignno}}
\def\eeq{\end{eqalignno}}
\def\eeq{\end{eqalignno}}
\def\bea{\begin{eqnarray}}
\def\eea{\end{eqnarray}}

\def\kt{\mbox{ $ k_T$}}
\def\xgam{\mbox{$x_\gamma$}}
\def\egamma{\mbox{$E_\gamma $}}
\def\f2gam{\mbox{$ F_2^\gamma $}}
\def\siggg{\mbox{$\sigma_{\gamma \gamma}^{\rm inel}$}}
\def\pghad{\mbox{$P_{\gamma}^{\rm had}$}}

\begin{document}
\begin{flushright}
LNF-08/022(P)\\
hep-ph/9807236\\
\end{flushright}
\begin{center}
{\large{\bf Eikonalised minijet model analysis of 
$\sigma_{\gamma \gamma}^{inel}$.}}
\\[1cm]
A. Corsetti $^a$, R.M. Godbole $^b$ \footnote{On leave of absence from
Dept. of Physics, Univ. of Bombay, India} and G. Pancheri $^c$. \\[1cm]
(a) Physics Department, Northeastern University, Boston, USA \\[0.5cm]
(b) Center for Theoretical Studies, Indian Institute of Science, 
Bangalore, India.\\[0.5cm]
(c) INFN - Laboratori Nazionali di Frascati, Frascati, Italy.
\end{center}

\begin{abstract}
We study the  theoretical predictions for the total inelastic
$\gamma \gamma$ cross-sections, with an emphasis on the eikonalised
minijet model (EMM). In the context of the EMM, we discuss a new
ansatz for the overlap function involving the photons. We
discuss the dependence of the EMM predictions on various input 
parameters as well as  predictions for \siggg\ from a simple
extension of the Regge Pomeron Exchange model. We then compare
both with the recent LEP data.

\end{abstract}

The measurement of the total photoproduction cross-sections at HERA
\cite {HERAZ,HERAH1} and the recent measurements of the hadronic
$\gamma \gamma $ cross-sections at LEP \cite{L3,OPAL}, have 
established that all total cross-sections, involving hadrons
 \cite{E710,CDF}
as well as photons,
% {\bf Will it be a good idea to refer to some 
%data for rising pp cross-sectiosn as well?, then we could 
%add afetr this comma a citation to all the 4 references
%HERAZ,HERAH1,LEP3,OPAL as well as either early ISR ref or recent UA1
%ref. what do u think?}
 rise as the c.m. energy of the colliding particles
increases.  The similarity in the energy dependence of {\it all}
 total cross-sections, suitably scaled to take into account the 
difference between hadrons and a photon, is striking.
\begin{figure}[hbt]
\leavevmode
\begin{center}
\mbox{\epsfig{file=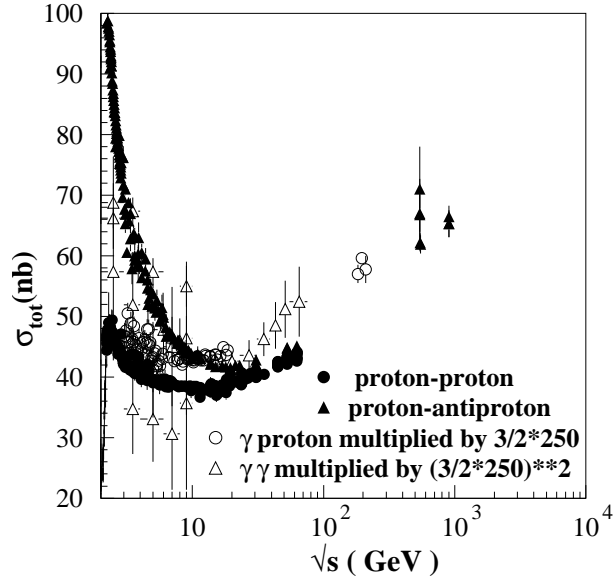,height=6cm}}
%pp_gamp_gamgamsec.eps,height=6cm}}
\vspace{1cm}
\end{center}
\caption{Energy dependence of $\sigma_{ab}^{\rm tot}$}
\label{fig0}
\end{figure}
Fig.(\ref{fig0}) demonstrates this. 
In the figure  the cross-sections for photon-induced processes
have been multiplied by  factors motivated by simple quark-model
considerations
and the probability for a photon to fluctuate in a $q \bar q$ pair
$P_{\gamma}^{\rm had}$, so as to facilitate comparison with $pp$
and $\bar p p$ data. We will discuss the choice of \pghad\ later.
In the light of the new $\gamma \gamma$ data the impetus to arrive 
at a unified description of the energy dependence of the total 
cross-sections, 
independent of whether the beam is a hadron or a photon, in terms of
a model based on perturbative QCD and the measured parton content
of the photon/hadrons, is now ever stronger.  The problem of hadronic
$\gamma
\gamma$ cross-sections at high energy \cite{SJOS,DESY} has an additional
significance in view of the large potential backgrounds  
that Beamstrahlung photons could cause at future Linear Colliders
\cite{messy}. Indeed this was one of the original motivations 
of this work. The other was to discuss in detail how different input parameters 
influence the models used to compare with the data. A preliminary 
version of our results along with the then available high energy data,
was presented at PHOTON-97~\cite{photon97}. It is evident from the
most recent literature on $\sigma^{\rm had}_{\gamma \gamma}$ that
this issue is still very much debated : the data from the OPAL collaboration
and L3 are consistent with each other within errors, but the models which 
describe the central values of one or the other of the two sets
may be rather different.
In this paper we shall compare the recent LEP data with the mini-jet
model,
trying to pinpoint  the  uncertainties of the model.
We shall present elsewhere the detailed analysis resulting from
varying all the parameters, here we shall limit ourselves to a summary of
such investigation.

The issue of energy dependence of the total cross-section
for high energy  particle scattering, is not a new one. As a 
matter of fact more than one model exist which try to explain
the observed energy dependence. One successful description of total
cross-sections is obtained in the Regge/Pomeron  model
\cite{DL}, in which t-channel exchanges in the elastic scattering
amplitude lead to the expression
$$\sigma^{tot}_{ab}=Y_{ab} s^{-\eta}+X_{ab}s^{\epsilon}$$
where $\eta$ and $\epsilon $  are related to the intercept at zero of the
leading Regge trajectory and of the Pomeron, respectively
 $\eta\approx 0.5$ and $\epsilon \approx 0.08$. This parametrization applies
successfully \cite{DL} to  photoproduction
and  is consistent within errors with  the lower
energy data on $\gamma \gamma$~\cite{SJOS,desy84,desy86,tpc,vepp4}. 
Since the lower energy data were characterized by large errors, 
this left wide margin \cite{LEP200,desy96} to predictions
about  where and by how much the cross-section rises.
% the when and how of the rise of the cross-section, 
However, they  provided the first
check of the hypothesis of factorization of the residues at the poles in the
Regge description of elastic and total cross-sections. In fact,
assuming the hypothesis of factorization, one can make a
prediction for 
$\gamma \gamma$ total  cross-section using
 $$Y_{ab}^2=Y_{aa} Y_{bb}\ \ \ \ \ \ X_{ab}^2=X_{aa} X_{bb}$$
and    extracting the coefficients X and Y from those for  the fit to
 photo-production and hadron-hadron data.

An alternative description of  total cross-sections  is given in terms
of  semi-hard QCD interactions, called minijet interactions, among
the partons of the colliding hadrons with partonic momenta of the
order of 1-2 GeV. 
It was proposed \cite{CLINE} 
that the rise of 
$\sigma^{\rm tot}$  with energy is driven by the increase with
energy of the number of hard interactions among the proton constituents,
i.e., in contemporary language, by the increase with energy
of the 
inclusive  productions for these minijets \cite{minijets}.
The lack of unitarity of the original model \cite{CLINE,minijets}
was cured by the introduction of the eikonalized mini-jet models
\cite{eikminijets}. These models have an immediate semi-classical derivation,
when applied to the semi-hard QCD contribution to $\sigma^{inel}$.
The eikonalized expression can in fact be obtained 
semi-classically by considering the number of collisions at fixed
impact parameter $b$. Under the assumption that the collisions are
independent of each other at any fixed value of b, one can assume that
 the distribution of r collisions around their average $n(b,s)$ follow
a Poisson distribution ${\cal P} (r,n)$ 
so that the sum over all possible collisions  
becomes \cite{treleani}
\be
\label{pois}
\sigma^{inel}_{QCD}=\int d^2{\vec b}\sum_{r=1}^{\infty} 
{\cal P} (r,n)= \int d^2{\vec b}\left[1-e^{-n(b,s)}\right]
\ee
Thus a constant average will give a constant cross-section, increasing 
number of
collisions will produce rising cross-sections.The QCD mini-jet model ascribes
to QCD the task of calculating the average number of semi-hard collision, 
identifying $n(b,s)$ with the total jet cross-section times a 
suitable function, which is responsible for the b-dependence. 
Since the jet cross-section depends sensitively on  the lowest 
transverse momentum, $p_{tmin}$, in the $p_t$-integration, 
the calculation is affected by the uncertainty due to the choice 
of $p_{tmin}$.
For purely hadronic scattering, the other large uncertainty
in the mini-jet model is the b-dependence of the partons in the proton, 
usually described through the Fourier transform of the electromagnetic form
factors of the hadrons. In order to describe the
rising proton-proton and proton-antiprotron cross-section,
and  to incorporate the fact 
that  the hadronic structure of all particles, photon included,
involves 
both a perturbative and nonperturbative part with an energy dependent 
relative weight,  the
mini-jet model has to be supplemented by the introduction of 
a non perturbative term in $n(b,s)$, thus 
introducing further parameters,
 like 
a different shape of the b-distribution as the energy increases or 
even the introduction of
  a sum of eikonalized
 functions \cite{SJOS,SARC} instead of a single one.
 The more the terms, the more the 
parameters that have to be introduced. Since 
 our aim here is to investigate the minimal parameter
dependence of the mini-jet model, we restrict ourselves to only one term.

Therefore, apart from  the assumption of one or more eikonals,
the  predictions of the eikonalised mini-jet model
in general will 
depend on 1)  the hard jet 
cross-section 
$\sigma_{jet}=\int_{p_{tmin}} {{d^2\hat{\sigma}}\over{dp_t^2}} dp_t^2$ 
which in turn depends on the minimum 
$p_t$ above which one can expect perturbative QCD to hold viz. $ p_{tmin}$,
and on the  parton densities in the colliding particles $a$ and $b$, 
2) the soft cross--section $\sigma^{\rm soft}_{ab}$ to be introduced to
describe the non perturbative region, 3) the overlap function
$ A_{ab}(b)$ in the $b-$integration.
 For photons,  the model  has apparently one more uncertainty, as it
has to incorporate \cite{ladinsky} the hadronisation probability 
 for the photon to fluctuate itself into a hadronic 
state,  $P^{had}_{\gamma}$. 
For photon induced processes then the eikonalisation leads to
\begin{equation}
\label{eikonal}
\sigma^{inel}_{ab} = P^{had}_{ab}\int d^2\vec{b}[1-e^{n(b,s)}],
\end{equation}
with the average number of collisions at a given impact
parameter $\vec{b}$ given by  
\begin{equation}
\label{av_n}
n(b,s)=A_{ab} (b) (\sigma^{\rm soft}_{ab} + {{1}\over{P^{\rm had}_{ab}}}
\sigma^{jet}_{ab})
\end{equation} 
where $P^{had}_{ab}$ is the probability that  the colliding particles
$a,b$ are both in a hadronic state,  
$A_{ab} (b)$ describes the transverse overlap of the  partons 
in the two particles  normalised to 1,
$\sigma^{\rm soft}_{ab}$ is the non-perturbative part of the cross-section
while $\sigma^{jet}_{ab} $ is the hard part of the cross--section (of order
$\alpha_{\rm em} $ or $\alpha_{\rm em}^2$ for 
$\gamma p$ and $\gamma \gamma$ respectively). 
Notice that, in the above definitions, $\sigma^{\rm soft}$
is a cross-section of hadronic size since the factor $P_{ab}^{had}$ has
already been factored out. 
Letting
\begin{equation}
\label{phad}
P_{\gamma p}^{had} = P_{\gamma}^{had}  \ \ \ and \ \ \ 
 P_{\gamma \gamma}^{had} \approx   (P_{\gamma}^{had})^2,
\end{equation}
one can   extrapolate the model from photoproduction to photon-photon 
collisions. Admittedly, this procedure is very simplistic,
as the probability $P_{\gamma}^{had}$ is certainly 
energy dependent. However, as in the case of the Regge-Pomeron model,
this approximation, i.e. eq.(\ref{phad}), allows for a check of factorization
ansatz. As for the energy dependence, a scaling property of
the eikonal model allows us to include it into $A_{ab}(b)$. Indeed,
by a simple change of variables,
it is easy to see \cite{review} that one can
 rewrite the above expression in such a way 
that mathematically the expression for the cross-section in
eq.(\ref{eikonal}) depends only on the combination $A_{ab}/P_{ab}^{\rm had}$.
This helps us reduce the number of parameters which
identify one given process and its energy dependence.
Thus the above scaling property
for $A_{ab}(b)/P_{ab}^{had}$ implies that 
we can fix a value for $P_{ab}^{had}$ at a given energy and vary only
$A_{ab}(b)$.

As we said, in order to clarify the limitations and parameter dependence of 
this model, we  restrict ourselves to a single eikonal. The hard jet
cross-sections are calculated in LO perturbative QCD and  use 
 photonic parton densities  GRV \cite{GRV} calculated to the
  leading order  as well as 
SaS densities, mode 1 \cite{SAS1}. We  
determine $\sigma_{\gamma \gamma}^{\rm soft}$ from 
$\sigma_{\gamma p}^{\rm soft}$ 
which in turn is  determined by a fit to the photoproduction data.
From inspection of the photoproduction  data, one
can assume that $\sigma^{\rm soft}$ should contain both a constant and a
term which falls with energy. Following the suggestion\cite{SARC}
\begin{equation}
\label{soft}
\sigma^{\rm soft}_{\gamma p} =\sigma^0_{\gamma p} +
{{{\cal A}_{\gamma p}}\over{\sqrt{s}}}+{{{\cal B}_{\gamma p}}\over{s}},
\end{equation}
we then calculate values for $\sigma^0_{\gamma p}, {\cal A}_{\gamma p}$ 
and ${\cal B}_{\gamma p}$ from a best fit \cite{thesis}
to the 
low energy photoproduction data, starting with the Quark Parton Model (QPM)
ansatz
$\sigma^0_{\gamma p}\approx {{2}\over{3}}\sigma^0_{pp}$,
where $\sigma^0_{pp}$ is the constant term in analogous eikonal type fits to
proton-proton scattering. For $\gamma \gamma$
collisions, we  shall repeat the QPM suggestion and propose
\begin{equation}
\sigma^{\rm soft}_{\gamma \gamma}={{2}\over{3}} \sigma^{\rm soft}_{\gamma p}.
\end{equation}

Most of the discussion about the mini-jet model predictions for the total
inelastic cross-section has so far
centered on the uncertainties generated by the choice of
  $p_{tmin}$ as far as the QCD calculation is concerned, and on
 $\sigma_{\rm soft}$ for the non-perturbative part, both in the case of 
photoproduction and $\gamma \gamma$ cross-sections
\cite{SJOS,SARC,FS}. On the other hand,  the effect of $A_{ab}(b)$ 
and $P^{had}_{ab}$ 
on the overall validity of this description has not really been assessed.
The real emphasis of our work is to discuss this point.
In this paper, we shall analyze the simplest and oldest  
ansatz concerning  the overlap function $A_{ab} (b)$
\begin{equation}
\label{aob}
A_{ab}(b)={{1}\over{(2\pi)^2}}\int d^2\vec{q}{\cal F}_a(q) {\cal F}_b(q) 
e^{i\vec{q}\cdot \vec{b}},
\end{equation}
 where ${\cal F}$ is the Fourier transform of the b-distribution
of partons in the colliding particles and can be
 obtained using for ${\cal F}$ the electromagnetic form 
factors of the colliding hadrons. 
For protons this is given by the dipole expression
\begin{equation}
\label{dipole}
{\cal F}_{prot}(q)=[{{\nu^2}\over{q^2+\nu^2}}]^2,
\end{equation}
with $\nu^2=0.71\ {\rm GeV}^2$. For photons a number of authors \cite{SARC,FLETCHER}, 
 on the basis of Vector Meson Dominance, have assumed the same functional 
form as for pion, i.e. the pole expression 
\begin{equation}
\label{pole}
{\cal F}_{pion}(q)={{k_0^2}\over{q^2+k_0^2}},
\end{equation}
with $k_0 = 0.735$ GeV from the measured pion form factors,
 changing the value of the scale parameter $k_0$, if necessary 
in order to fit the data. 
We would like to adopt here a different philosophy, i.e. that the b-space 
distribution of partons is the Fourier transform of the
transverse momentum distribution of the colliding system~\cite{BN}. 
To leading order,   this transverse momentum
distribution can be  entirely due to an intrinsic
transverse momentum of partons in the parent hadron, but while
the intrinsic transverse momentum ($k_T$) distribution of partons in a proton 
is normally taken to be  Gaussian,  a choice which can be justified in QCD
based models \cite{nak},  in the case 
of photon  the origin of all  partons can, in principle, be traced back to 
the hard vertex $\gamma \ q \bar q$.  Therefore, also in the intrinsic
transverse momentum
philosophy,  one can expect the  
$k_T$ distribution 
of photonic partons to be different from that of the partons in the proton. 
The expected functional dependence can be deduced  using the origin 
of photonic partons from the $\gamma \rightarrow q \bar q$ splitting. We
present below a discussion of the same and then proceed to assess the
effect of this ansatz for $A_{ab}(b)$. 

To do this consider a single resolved diagram (say) for $\gamma \gamma 
\rightarrow $ 2 hard jets given below in 
\begin{figure}
\leavevmode
\begin{center}
\mbox{\epsfig{file=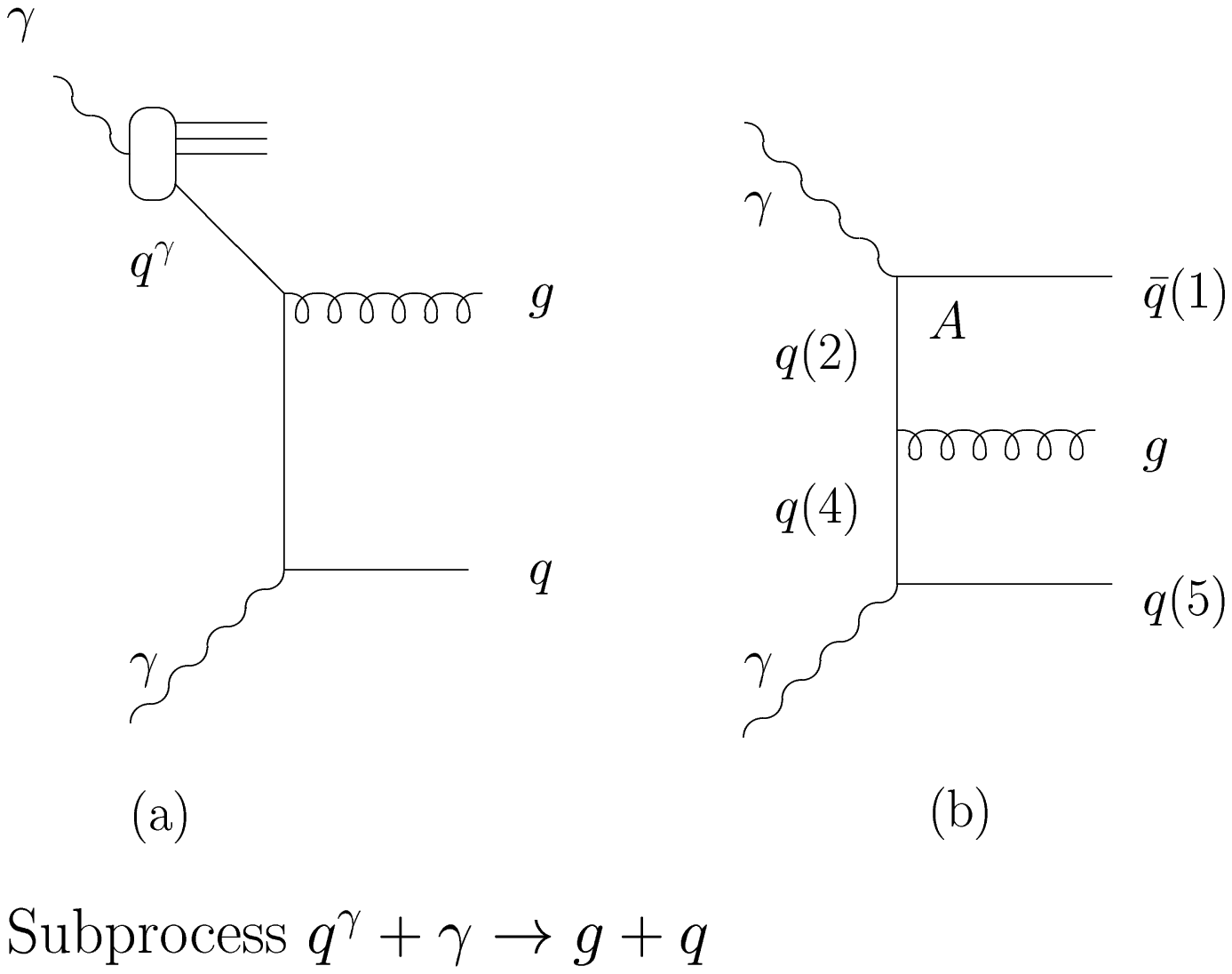,height=6cm}}
\end{center}
\caption{A single resolved process and corresponding perturbative diagram.}
\label{fig1}
\end{figure}
Fig.(\ref{fig1}.a).
Perturbatively the same diagram can be drawn as shown in 
Fig.(\ref{fig1}.b). The quark (2) in this figure is `almost' on shell 
and that is why it can be looked upon as $q^\gamma$ in Fig.(\ref{fig1}.a).  
 Due to the hard $ \gamma \bar q (1) q(2) $ vertex at `A' in  Fig.(\ref{fig1}.b) 
$\bar q (1) $ and hence $q (2)$ can appear
with sizable transverse momentum. Hence in the structure function language 
one can talk of an intrinsic transverse momentum ($k_T$) of the photonic 
partons. Assuming $\bar q (1)$ to be on shell the 4-mmta. of the three 
particles $\gamma , \bar q (1)$ and $q (2)$ are
\beq\label{momenta}
\gamma & = E_\gamma (1,0,0,1); \\
1 & = (\egamma (1-\xgam),0,\kt,\sqrt {\egamma^2 (1-\xgam)^2 - k_T^2} );\\
2 & = (\egamma \xgam , 0, -\kt, \egamma - \sqrt{\egamma^2 (1-\xgam)^2 - 
k_T^2} ).
\eeq

\begin{figure}[hbt]
\leavevmode
\begin{center}
\mbox{\epsfig{file=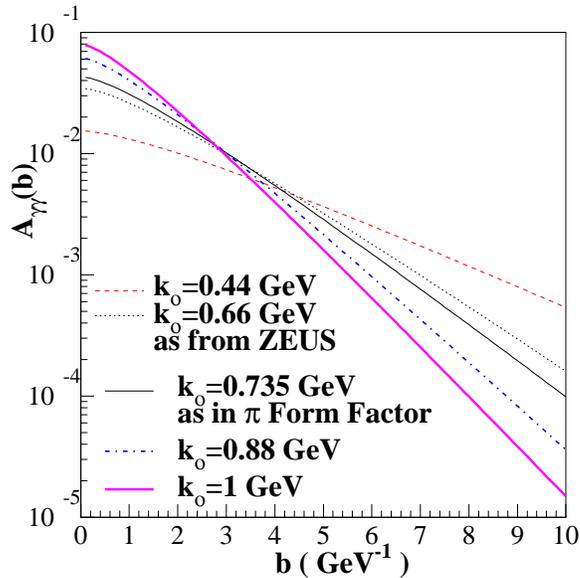,height=6cm}}
%aobgamgam.eps,height=6cm}}
\vspace{1cm}
\end{center}
\caption{The overlap function for 
 $\gamma \gamma$ collisions according to the two different models discussed 
in the text. The value $k_0=0.735~$ GeV corresponds to using the pion form 
factor for the photon overlap function. The other values correspond to 
the experimental value given by the ZEUS Collaboration: 
$0.66 \pm 0.22$ GeV. The value of 1 GeV corresponds approximately
 to the maximum value at $90 \%$ confidence level.}
\label{aobf}
\end{figure}

The virtuality of parton '2' is
\be
(2)^2 \equiv t(2) = 2 \egamma ^2  (1 - \xgam) \left[ -1 + 
\sqrt{1 - {k_T^2 \over {\egamma^2 (1 - \xgam)^2}}} \right].
\ee
The structure function language makes sense when \kt\
 is small. Then expanding the root we get 
 \be
 t(2) = - {k_T^2 \over (1 - \xgam)}.
 \ee
 The dominant part of the perturbative diagram  in Fig.(\ref{fig1}.b)
 which is approximated by the single resolved diagram in the structure 
 function language, is given by small values of $t(2)$, as the cross-section  
 (in the leading log approximation) is
 \be
 {d \sigma \over dt(2)} \propto {1 \over t (2) }
 \ee
 This immediately tells us that 
 \be
 {d q^\gamma \over dk_T^2 } \propto {1 \over k_T^2}
 \ee
 Of course the distribution has to be regularised. One expects the
 nonperturbative effects to keep quark (2) always (slightly) off mass
 shell. Phenomenolgically this would imply an intrinsic \kt\
 distribution given by
 $$
 f(\kt) = {C \over (k_T^2 + k_{0}^2)}
 $$
Finally, we have to choose the normalisation $C$ such that the eventual
$A_{ab}(b)$ satisfies 
$$
\int d^2{\vec b} A_{ab}(\vec b) = 1.
$$
and this gives 
$$
f(k_T) = {1 \over 2 \pi} {k^2_{0} \over (k^2_T + k^2_{0})} . 
$$
To summarize, while for the proton the transverse momentum
distribution model for $A_{ab}(b)$ would correspond to use of a Gaussian 
distribution instead of the dipole expression of eq.(\ref{dipole}),
for the photon one can argue that the intrinsic transverse momentum ans\"atze
\cite{rohini} would imply the use of a different value of the parameter 
$k_0$, which is extracted from data involving `resolved'
photon interactions \cite{ZEUS}, in the pole expression for the form factor.
By varying $k_0$ one can then explore  various
possibilities, i.e. the VMD/pion hypothesis if $k_{0}=0.735 $ GeV,
or the intrinsic transverse distribution case for other values 
of $k_0$. Still another possibility, not in
contradiction with the above, is that $A_{ab}(b)$ is the Fourier transform of
the transverse momentum distribution of the initial collinear
parton pair, to be evaluated using soft gluon summation techniques
\cite{BN}. However, in this letter we limit ourselves to models with the 
intrinsic transverse momentum ansatz for the photon.
Assuming the functional expression described above, the overlap
function for $A_{ab}(b)$ becomes
\be
\label{aobgg}
A_{\gamma \gamma}(b)={{1}\over{4\pi}}k_0^3b {\cal K}_1(bk_0)
\ee 
  To show the dependence of 
$A_{\gamma \gamma}(b)$ from the scale parameter $k_0$ (which in principle
could be energy dependent, we stress again), we plot
in Fig.(\ref{aobf}) 
 the funtion $A_{ab}(b)$ for $\gamma \gamma$ scattering for various values of
 $k_0$. Notice that the  
region most important to this calculation is for large values of the
parameter b, after  the overlap function changes trend, and where fall 
with b is faster for larger $k_0$ values. This is in agreement with 
the intuitive idea that higher $k_0$ values correspond to
hardening of the scale of the subprocesses.
  
As for $P^{had}_\gamma$,  which is clearly expected to be 
${\cal O} (\alpha_{\rm em})$, VMD prescriptions would 
suggest 
\begin{equation}
\label{PVMD}
P_\gamma^{had} = P_{VMD}=\sum_{V=\rho,\omega,\phi} {{4\pi 
\alpha_{\rm em}}
\over{f^2_V}}\approx {{1}\over{250}}
\end{equation}
We shall fix the value in such a way as to obtain a good fit to the
photoproduction data and then use factorization for the comparison
with the $\gamma \gamma$ cross-section. 
This corresponds to a value \cite{FLETCHER} of 
  1/204, which includes a non-VMD contribution of $\approx 20\%$.
 We shall not discuss this point at length, since  
 for any given  value, $P_{had}$ can be absorbed into a 
  redefinition of the scale parameters $k_0$ and $\nu$ through a simple change
 of variables
  \cite{review}. If $P^{had}_{\gamma}$ were to be energy dependent, this would
then result into energy dependence of the scale parameters.

After this rather general introduction, whose aim was to
 establish the range of variability of the quantities 
involved in the mini-jet calculation of
 total inelastic photonic cross sections, we shall now present 
 the predictions of the eikonalized minijet model with respect to the
 presently available data for photon-photon scattering.

  We start with the photoproduction cross-section from HERA,
using  GRV (LO)  densities and different values of $p_{tmin}$
for the jet cross-section. For $\sigma_{\rm soft}$ we proceed as
described, obtaining two good fits. We choose the set
\begin{equation}
\sigma^0_{\gamma p}=31.2~{\rm mb}, {\cal A}_{\gamma p}=0.0~{\rm mb}~,
{\cal B}_{\gamma p}= 63.1~{\rm mb ~GeV}^2
\end{equation}
\begin{figure}[htb]
\leavevmode
\begin{center}
\mbox{\epsfig{file=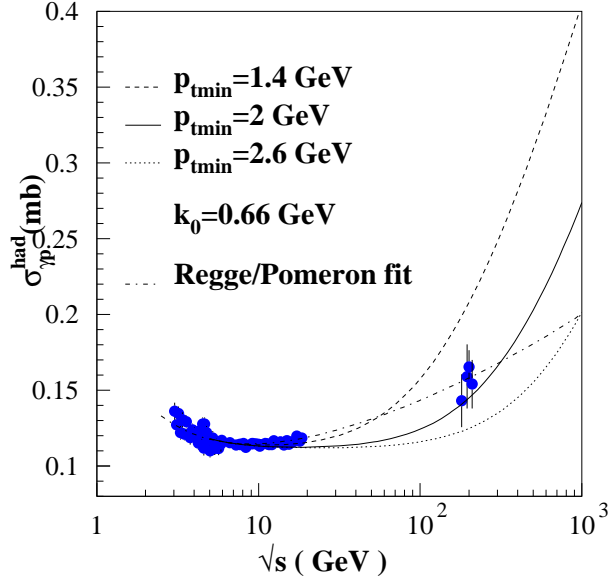,height=6cm}}
\vspace{1cm}
\end{center}
\caption{Total inelastic photon-proton cross-section
 for GRV densities and various $p_{tmin}$ values,
 compared with data and Regge/Pomeron parametrization \protect\cite{DL}}
\label{gamprot}
\end{figure}
 We show  the mini-jet result in Fig.(\ref{gamprot}),
 using the form factor model for $A_{\gamma p}(b)$,
i.e. eq.(\ref{aob}) with $k_0$ modified so as to take into account the 
intrinsic transverse momentum hypothesis for the photon and the findings
 by the ZEUS Collaboration \cite{ZEUS}, i.e. $k_0=0.66 \pm 0.22\ $GeV.
  The curves  are obtained summing both the resolved and the direct
 contribution which is order $\alpha_{\rm em}$ relative to the hadronic term.
We observe that it is very difficult with a single
eikonal, the same $A_{\gamma p}(b)$ and the same set of parton densities,
 GRV  for
 the
proton as well as for the photon, to fit both the
beginning of the rise of the cross-section and 
 the high energy points. If one chooses to fit the high energy rise, then
a good choice could be $p_{tmin}=2\ $GeV, but the low energy 
region would be better described  by a smaller $p_{tmin}$, like
 $p_{tmin}\le 1.4\ $ GeV. This is the most difficult point of
the mini-jet model, i.e. whether it is possible to understand both the
beginning of the rise as well the
rise in the high energy region. Whether this dificulty has to be
ascribed to our still rough understanding of the impact parameter description
or to non perturbative effects in the transition between
$\sigma^{soft}$ and $\sigma^{jet}$ or to a combination of both these effects,
will be investigated in future papers. Here, for a comparison, we show in Fig.(\ref{gamprot})
the fit from the Regge/Pomeron exchange model \cite{DL}.

We now apply the  criteria and parameter set used in
 $\gamma p$ collisions to the case of photon-photon collisions,
i.e. $P_{\gamma}^{had}=1/204$, $p_{tmin}=2\ $ GeV, 
$A_{\gamma \gamma}(b)$ from eq.(\ref{aob}), and the value $k_0=0.66\ $ GeV for
the photon scale parameter.

\begin{figure}[hbt]
\leavevmode
\begin{center}
\mbox{\epsfig{file=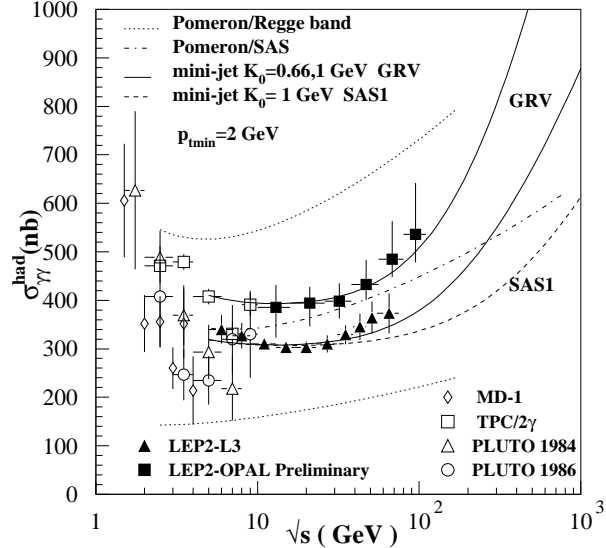,height=6cm}}
\vspace{1cm}
\end{center}
\caption{Total inelastic photon-photon cross-section in the eikonalized
 mini-jet model with  $p_{tmin}=2\ $ GeV,
 compared with data \protect\cite{L3,OPAL,desy84,desy86,tpc,vepp4}
 and Regge/Pomeron parametrization (see text). L3 data do not
include the quoted \protect\cite{L3} uncertainties on scale 
and overall normalization.
 The two
lower mini-jet curves correspond to $k_0=1 \ $ GeV with GRV and SAS1 densities.
The highest one is   for GRV densities and $k_0=0.66 $ GeV.}
\label{gamgam}
\end{figure}

We show the results in Fig.(\ref{gamgam}). 
 Clearly, all the quantities
involved
in photoproduction appear here squared, so that the possibilty of measuring
 the photon
related parameters in photon-photon scattering is enhanced by
 the sensitivity of
the cross-section to these parameters. 
To show the dependence of the model upon the uncertainties we have just 
discussed, we have obtained also fits varying the  parton densities
 and/or  $k_0$. In particular,
we can check the intrinsic transverse momentum ans\"atze through 
variations in $k_0$ values. Fits obtained with 
a value of 
$k_0$, of order 1 GeV, in keeping with the hypothesis of a harder
scale are also shown. For this value, we plot the result with
both GRV and SAS1 parton densities. The highest of the two
full  lines corresponds exactly to the
 same parameter set used in the photoproduction case, Fig.(\ref{gamprot}),
and appears, in this model, to be in good agreement with the preliminary
 results from the
OPAL \cite{OPAL} Collaboration, whereas the L3 results, everything else being
the same, would favour a higher
$k_0$ value. Notice however that if one include the errors due to normalization
and scale to the ones explicitly indicated
 in Table 2
of ref. (\cite{L3}), then the two sets of data are consistent with each 
other within one
standard deviation.

All the predictions of the minijet model are also compared
with 
predictions (Pomeron/SaS) 
based on a Pomeron/Regge type parametrization\cite{SJOS}, using factorization
of the residues as described before. and 
 using for
$\eta$ and $\epsilon$ the average values from the Particle Data 
Group\cite{PDG},
and for X and Y the average between
 $pp$ and $p {\bar p}$. Finally, the dashed
band  has been obtained 
 using the errors on X and Y, whereupon
the large errors on Y coming from photoproduction are responsible
 for the large band shown. One notices the same occurrence as in
the photoproduction data, i.e. that the
curvature with which the
cross-sections rise is quite different in mini-jet and Pomeron/Regge models,
 as observed already 
in all total cross-sections. 

In conclusion, we have compared recent data on photon-photon total
 cross-section with predictions from both a Regge/Pomeron type parametrization 
 as well as from the mini-jet models. We stress the fact that our mini-jet 
 analysis is based on currently used QCD parton densities and on a clear 
 specification of the variable parameters used. We  see here that the data are
 well described by the minijet model where the same set of 
parameters affords a acceptable description of both , the photoproduction as
 well as $\gamma \gamma$
 inelastic cross-sections.


\begin{thebibliography}{99}
\bibitem{HERAZ} Zeus Collaboration, Phys. Lett. {\bf B 293} (1992), 465; 
Zeit. Phys. {\bf C 63} (1994) 391.
\bibitem{HERAH1} H1 Collaboration, Phys. Lett. {\bf B 299} (1993) 374;
Zeit. Phys. {\bf C69} (1995) 27.
\bibitem{L3} L3 Collaboration, Phys. Lett. {\bf B 408} (1997) 450.
\label{L3}
\bibitem{OPAL} OPAL Collaboration :  F.W\"ackerle, to be published in the
Proceedings of   XXVII International Symposium on Multiparticle 
Dynamics, Frascati 8-12 September 1997, Conference
Suppl. Nucl. Phys. B, Eds. G. Capon et al. and
OPAL Physics Note 320, september 9, 1997. 
\label{OPAL}
\bibitem{E710} E. Amos et al., Phys. Rev. Lett. {\bf 63} (1989) 2784.
\label{E710}
\bibitem{CDF} CDF Collaboration, Phys. Rev.  {\bf D 50} (1994), 5550.
\label{CDF}
\bibitem{SJOS}
G. Schuler and T. Sj\"ostrand, Phys. Lett. {\bf B 300} (1993) 169, 
 Nucl. Phys. B407  (1993) 539.
\bibitem{DESY} E. Accomando et al., {\it Physics with Linear Colliders},
DESY 97-100, to be published in Physics Report. See $\gamma \gamma$ section 
and references therein. 
\bibitem{messy}
M. Drees and R.M. Godbole, Phys. Rev. Lett. {\bf 67} (1991) 1189,
P. Chen, T.L.. Barklow and M. E. Peskin, Phys. Rev. {\bf D 49} (1994) 3209.
\label{messy}
\bibitem{photon97}
A. Corsetti, R.M. Godbole and G. Pancheri, in Proccedings of {\it
PHOTON' 97}, Eds. A. Buijs and F.C. Erne, Egmond aan Zee, May 1997,
World Scientific, {\bf hep-ph/9707360}.
\bibitem{DL} A. Donnachie and P.V. Landshoff, Phys. Lett. {\bf B 296} (1992) 227.
\label{DL}
\bibitem{desy84}Ch. Berger et al., PLUTO Coll., Phys. Lett.  {\bf B 149} (1984) 421; 
 Zeit Phys. {\bf C 26} (1984) 353.
\bibitem{desy86} M. Feindt, {\it Recent PLUTO Results on Photon Photon
Reactions}, Proceedings of the 1986 Paris Workshop on Photon-Photon Collisions,
Eds. A. Courau and P. Kessler,  World Scientific.
\bibitem{tpc}
H. Aihara et al., TPC/2$\gamma$ Coll., Phys. Rev. {\bf D 41} (1990) 2667
\bibitem{vepp4} S.E. Baru et al., MD-1 Coll., Zeit Phys. {\bf C 53} (1992) 219.
\bibitem{LEP200} P. Aurenche et al, {\it $\gamma \gamma$ Physics at LEP2},
 hep-ph/9601317, pub. in Proceedings of the Workshop on Physics at LEP2, 
CERN 96-01, v.1.
\bibitem{desy96}A. Corsetti, R. Godbole and G. Pancheri, in
Proceedings of the Workshop on $e^+e^- $ Collisions at TeV Energies, 
Annecy, Gran Sasso, Hamburg, DESY 96-123D, June 1996, page. 495.
\label{desy96}

%\bibitem{PW} M.R. Pennington and M. R. Whalley, Journ. Phys. G 20 Suppl.
% 8A (1994) A1-A417.
%\label{PW}
%\bibitem{PDG}
% Particle Data Group, Physical Review D54 (1996) 191.
%\label{PDG}
\bibitem{CLINE}
D.Cline, F.Halzen and J. Luthe, Phys. Rev. Lett. {\bf 31} (1973) 491.
\label{CLINE}
\bibitem{minijets}
A. Capella and J. Tran Thanh Van, Z. Phys. {\bf C 23} (1984)168.
G. Pancheri and C. Rubbia, Nuclear Physics {\bf A 418} (1984) 117c.
T.Gaisser and F.Halzen, Phys. Rev. Lett. {\bf 54}  (1985) 1754.
P. l`Heureux, B. Margolis and P. Valin, Phys. Rev. {\bf D 32} (1985) 1681.
G.Pancheri and Y.N.Srivastava,  Physics Letters {\bf B 158} (1986) 402.
\label{minijets}
\bibitem{eikminijets}
L. Durand and H. Pi, Phys. Rev. Lett. {\bf 58} (1987) 58.
A. Capella, J. Kwiecinsky, J. Tran Thanh, Phys.\ Rev.\ Lett.\ 58 (1987) 2015.
M.M. Block, F. Halzen, B. Margolis, Phys. Rev. {\bf  D 45} (1992) 839.
\label{eikminijets}
\bibitem{treleani}
 D. Treleani and L. Ametller, Int.\ Jou.\ Mod.\ Phys.\ {\bf A 3} (1988) 521 
\label{treleani}
\bibitem{ladinsky}
J.C. Collins and G.A. Ladinsky, Phys. Rev. {\bf D 43} (1991) 2847.
\label{ladinsky}
\bibitem{review}
M. Drees, Univ. Wisconsin report {\bf MAD/PH-95-867},{\it  Proceedings of the
4th workshop on TRISTAN physics at High Luminosities}, 
KEK, Tsukuba, Japan, Nov. 1994;
M. Drees and R. Godbole, J. Phys. {\bf G 21} (1995) 1559.
\label{review}
\bibitem{GRV}
M. Gl\"uck, E. Reya and A. Vogt, Phys. Rev. {\bf D 46} (1992) 1973.
\label{GRV}
\bibitem{SAS1} G. Shuler and T. Sjostrand, Zeit. Phys. {\bf C 68} (1995) 607;
 Phys., Lett. {\bf B 376} (1996) 193.
\bibitem{SARC}
K. Honjo,  L. Durand, R. Gandhi, H. Pi and I. Sarcevic, Phys. Rev. {\bf D 48} 
(1993) 1048. 
\label{SARC}
\bibitem{thesis} A. Corsetti, September 1994 Laurea Thesis, University of Rome
La Sapienza.
\label{thesis}
\bibitem{FS}
J.R. Foreshaw and J.K. Storrow, Phys. Lett. {\bf B 278} (1992) 193;
 Phys. Rev.{\bf D 46} (1992) 3279.
\label{FS} 
\bibitem{FLETCHER}
R.S. Fletcher , T.K. Gaisser and F.Halzen, Phys. Rev. {\bf D 45} (1992) 377; 
erratum Phys. Rev. {\bf D 45} (1992) 3279.
\label{FLETCHER}
\bibitem{BN} A. Corsetti,  Grau, G. Pancheri and Y.N. Srivastava, Phys. Lett.
{\bf B 382}(1996) 282.
\label{BN}
%\bibitem{DG}
%M. Drees and K. Grassie, Z. Phys. {\bf C 28} (1985) 451.
%\label{DG}
\bibitem{nak} A. Nakamura, G. Pancheri and Y.N. Srivastava, Zeit. Phys.
{\bf C 21} (1984) 243.
\label{nak}
\bibitem{rohini} J. Field, E. Pietarinen and K. Kajantie, Nucl. Phys. 
{\bf B 171} (1980) 377; M. Drees, {\it Proceedings of 23rd International 
Symposium on Multiparticle Dynamics}, Aspen, Colo., Sep. 1993. Eds. M.M. Block
and A.R. White. 
\bibitem{ZEUS}
M. Derrick et al., ZEUS coll., Phys. Lett. {\bf B 354} (1995) 163.
\label{ZEUS}
\bibitem{PDG}
 Particle Data Group, Physical Review D54 (1996) 191.
\label{PDG}
%\bibitem{DGMINI}
%M.Drees on minijets
%\label{DGMINI}
%these reference pertain to A(b)
%\bibitem{GODBOLE}
%on intrinsic transverse momentum of photons
%\label{GODBOLE}
%the following references pertain to minijet cross-sections

%the next references pertain to Phadronic
\end{thebibliography}
\end{document}